\documentclass[
    ,final            
  ]
  {aipproc}

\layoutstyle{6x9}

\newcommand\lsim{\mathrel{\rlap{\lower4pt\hbox{\hskip1pt$\sim$}}
        \raise1pt\hbox{$<$}}}
\newcommand\gsim{\mathrel{\rlap{\lower4pt\hbox{\hskip1pt$\sim$}}
        \raise1pt\hbox{$>$}}}
\newcommand{\msun}{{\rm M_{\odot}}}
\newcommand{\apj}{ApJ}
\newcommand{\prd}{Phys. Rev. D}

\newcommand{\apjl}{ApJL}
\newcommand{\mnras}{MNRAS}

\newcommand{\Msol}{M_{\odot}}

\newcommand{\Mpc}{\;\mathrm{Mpc}}

\begin{document}

\title{The Origin and Detection of High-Redshift Supermassive Black Holes}

%


\author{Zolt\'an Haiman}{
  address={Department of Astronomy, Columbia University\\550 West 120th Street, New York, NY 10027, USA}
}

\begin{abstract}
  Supermassive black holes (SMBHs) are common in local galactic
  nuclei, and SMBHs as massive as several billion solar masses already
  exist at redshift $z=6$. These earliest SMBHs may arise by the
  combination of Eddington-limited growth and mergers of stellar-mass
  seed BHs left behind by the first generation of metal-free stars,
  or by the rapid direct collapse of gas in rare special environments
  where the gas can avoid fragmenting into stars. In this
  contribution, I review these two competing scenarios. I also briefly
  mention some more exotic ideas and how the different models may be
  distinguished in the future by {\it LISA} and other instruments.
\end{abstract}

\maketitle


\section{Introduction}

The discovery of very bright quasars, with luminosities $\ge
10^{47}~{\rm erg~s^{-1}}$, at redshift $z\simeq 6$ in the Sloan
Digital Sky Survey (SDSS) suggests that some SMBHs as massive as a
few$\times10^9~\msun$ already existed when the universe was less than
1 Gyr old (see, e.g., ref.~\cite{Fanreview06} for a review).  The
presence of these SMBHs presents a puzzle.  Metal--free stars, with
masses $\sim 100~\msun$, are expected to form at redshifts as high as
$z\gsim 25$ \cite{ABN02,BCL02,YOH08}, and leave behind remnant BHs
with similar masses \cite{Heger+03}.  However, the natural time-scale,
i.e. the Eddington time, for growing these seed BHs by $\gsim 7$
orders of magnitude in mass is comparable to the age of the universe
(e.g. ref.\cite{HL01}). This makes it difficult to reach $10^9~\msun$
without a phase of rapid (at least modestly super--Eddington)
accretion, unless a list of optimistic assumptions are made in
hierarchical merger models, in which multiple seed BHs are allowed to
grow without interruption, and to combine into a single SMBH
\cite{Haiman04,YM04,BSF04,Shapiro05,VR06,PDC07,Li+07,SSH09,TH09}.

An alternative class of explanations involves yet more rapid gas
accretion or collapse
\cite{OH02,BL03,KBD04,LN06,SS06,BVR06,VLN08,WA08,RH09b,SSG10,SBH10}. In
this family of models, primordial gas collapses rapidly into a SMBH as
massive as $10^4-10^6~\msun$, either directly, or possibly by
accreting onto a pre--existing smaller seed BH \cite{VR05}, or going
through the intermediate state of a very massive star \cite{BL03}, a
dense stellar cluster \cite{OSH08,DV09}, or a ``quasistar''
\cite{BRA08}.  These so--called ``direct collapse'' models involve
metal--free gas in relatively massive ($\gsim 10^8~\msun$) dark matter
halos at redshift $z\gsim 10$, with virial temperatures $T_{\rm
vir}\gsim 10^4$K.  The gas that cools and collapses in these halos
must avoid fragmentation, shed angular momentum efficiently, and
collapse rapidly.


\section{Growth from Stellar-Mass Seeds}
\label{sec:stellarseed}

Several authors have worked out the growth of SMBHs from stellar--mass
seeds, by following the build-up of dark matter (DM) halos, and using
simple prescriptions to track the formation of seed BHs, their
subsequent growth by accretion, and their mergers.  This can be done
either semi--analytically \cite{HL01,WL03,Haiman04,Shapiro05}, using
Monte-Carlo realizations of the DM merger trees
\cite{YM04,BSF04,VR06,TH09}, or based on cosmological hydrodynamics
simulations \cite{Li+07,PDC07,SSH09}.

The uncertainties about the statistics of the DM halo merger trees are
essentially negligible, since DM halo formation has been directly
resolved in numerical simulations at the relevant low masses (down to
$\sim 10^6~{\rm M_\odot}$) and high redshifts (out to $z\approx
30$). The accuracy of the merger trees is limited mainly by the
$5-10\%$ uncertainty in the normalization of the primordial power
spectrum, $\sigma_{8h^{-1}}$, and by the need to extrapolate the
primordial power spectrum 2-3 orders of magnitude below the spatial
scales on which it has been directly constrained.\footnote{In models
in which the small-scale power is suppressed, such as warm dark
matter, this extrapolation can dramatically reduce the number of
high-redshift halos, making it much harder to form the seeds of the
$z=6$ SMBHs~\cite{BHO01}.}

The most important -- and still rather uncertain -- ingredients of
this 'stellar seed' scenario can be summarized as follows. (i) What is
the threshold mass (or virial temperature, $T_{\rm seed}$) for early
DM halos in which Pop III stars can form? A reasonable guess is $T_{\rm
seed}=$ few $\times$ 100 K, which allows molecular ${\rm
H_2}$--cooling \cite{HTL96,Tegmark+97}. (ii) In what fraction ($f_{\rm
seed}$) of these halos do seed BHs form? This is a more difficult
question, since various feedback processes (due to radiation, metal
pollution, or mechanical energy deposition) could suppress Pop III star
formation in the vast majority of early low--mass halos. The answer
also depends on the IMF of Pop III stars, since whether the stars leave
a BH remnant or explode as pair-instability SNe depends on their
masses.  (iii) What is the time--averaged accretion rate of the seed
BHs?  This is conveniently parameterized by a duty cycle $f_{\rm
duty}$, defined as the fraction of the mass accretion rate that would
produce the Eddington luminosity, if $\epsilon\approx 10\%$ of the
rest mass was converted to radiation (so that $f_{\rm duty}=1$ is the
fiducial Eddington rate).  The expectation is that $f_{\rm duty}$ is
less than unity due to radiative feedback (but in practice, if the
accretion is radiatively inefficient, or if the radiation is trapped
or is beamed and ``leaks out'', then $f_{\rm duty}$ could exceed
unity).  (iv) Finally, what happens when DM halos merge? The simplest
and most optimistic assumption is that the BHs to promptly coalesce,
as well.  However, even if dynamical friction on the DM is efficient,
it is possible that, due to the radiation of its parent star, the
remnant BHs are no longer embedded in dense enough gas to allow
this. Furthermore, even if the BHs coalesce, the merged binary BH can
be ejected from the shallow potential wells ($\sim 1 $km/s) of the
early halos by the gravitational ``kick'', and effectively lost.  This
depends on the recoil speed, which depends strongly on the mass ratio
and on the spin vectors of the two BHs.

\begin{figure}
  \includegraphics[height=.3\textheight]{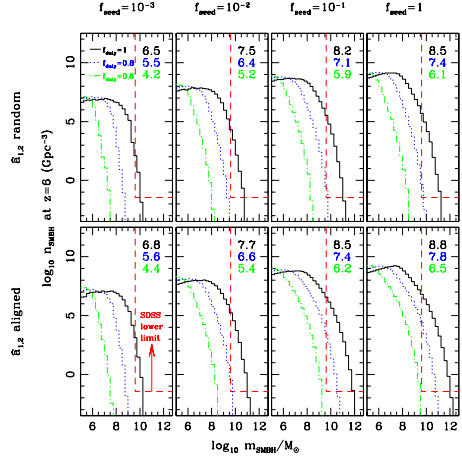}
  \caption{The comoving number densities of SMBHs in different mass
    bins at redshift $z=6$.  The 24 different models shown in the
    figure assume different parameter combinations as follows. The
    columns, from left to right, adopt $f_{\rm seed}=10^{-3}$,
    $10^{-2}$, $10^{-1}$, $1$.  The top row assumes a random binary
    spin orientation, and the bottom row assumes that BH spins are
    aligned with the binary's orbital angular momentum.  In each
    panel, the time--averaged mass--accretion rates, in Eddington
    units, are assumed to be either $f_{\rm duty}=1$ (black solid
    curves), $f_{\rm duty}=0.8$ (blue dotted), and $f_{\rm duty}=0.6$
    (green dash-dotted).  The numbers in the upper-right corners
    represent the total mass density in SMBHs
    $\log_{10}[\rho_{\bullet}/(\Msol \, \Mpc^{-3})]$ for each model.  The
    red dashed line demarcates the abundance of $z\approx 6$ SMBHs
    with $m\gsim 10^{9.6}\Msol$ already observed in the SDSS.}
\label{fig:SMBHmassfunction}
\end{figure}

In Figure~\ref{fig:SMBHmassfunction}, we show SMBH mass functions at
$z=6$, illustrating the impact of the above assumptions, taken from a
recent example of the Monte Carlo merger tree approach \cite{TH09}.
The mass functions were constructed from the merger histories of
$\approx10^5$ DM halos with masses $M>10^{8}\Msol$ at redshift $z=6$.
A robust conclusion for a model to produce enough $z=6$ SMBHs is that
$f_{\rm duty}\gsim 0.6$ -- namely the $\approx 100~{\rm M_\odot}$
stellar seed BHs must accrete near the Eddington rate nearly all the
time.  The initial BH occupation fraction also has to be $f_{\rm
seed}\gsim 10^{-3}$.  Finally, if the initial seeds are rare ($f_{\rm
seed}=10^{-3}-10^{-2}$), then gravitational kicks do {\em not} have a
big impact, and it makes little difference to the SMBH mass function
whether spins are aligned or randomly oriented. This is because in
this case, the few ``lucky'' seeds that form earliest already have a
chance to grow by $\gsim$ two orders of magnitude in mass before
encountering their first merger.  The masses of the two BHs at the
merger are then very unequal ($q=M_1/M_2\lsim 0.01$), making kick
velocities too low ($\sim 1$km/s; irrespective of the spins) to lead
to ejection.

An important additional issue is that in those models that satisfy the
SDSS constraint on the SMBH abundance (upper right corners in
Figure~\ref{fig:SMBHmassfunction}, marked in red), the massive end of
the SMBH mass function is extremely steep.  This prediction is not
surprising, as the most massive SMBHs reside in few $\times10^{12}{\rm
M_\odot}$ halos, which probe the $5\sigma$ tail of the halo mass
function at $z=6$ (and there are indeed $\approx 10^8$ times as many
few$\times10^{9}{\rm M_\odot}$ halos, which host $\sim 10^6{\rm
M_\odot}$ BHs).  It does mean, however, that the total mass density in
SMBHs with masses above $\gsim 10^5{\rm M_\odot}$ BHs (shown by the
numbers in the upper right corners in
Figure~\ref{fig:SMBHmassfunction}) are overpredicted by a factor of
$10^2-10^3$. The mass density of such SMBHs at $z\approx 0$ is
inferred to be several$\times 10^5{\rm M_\odot Mpc^{-3}}$, and the
expectation is that most ($\gsim 90\%$) of this mass was accreted well
after $z=6$ \cite{Shankarreview}.  Some strong feedback is therefore
needed to eliminate this significant overprediction. Possible
candidates for this are radiative feedback internal to halos, which
maintains the ``$M-\sigma$ relation'' in ultra--high redshift, low
mass halos, or the termination of Pop III star formation, at redshifts
as high as $z\sim 20$, due to Lyman Werner radiation or metal
pollution.

Finally, it is worth emphasizing that the mass accretion rate
corresponding to the Eddington limit -- for the fiducial radiative
efficiency of $\epsilon\equiv L/\dot{m}c^2=0.1$ for converting mass to
radiation -- would need to be exceeded only by a factor of a $\sim$ few
to make the growth from stellar seeds much easier.  Radiative feedback
is usually expected to lead to sub--Eddington rates
(e.g. \cite{AWA09}), and in spherical symmetry, the accretion was
recently to shown to be episodic, with $f_{\rm duty}\approx 0.3$
\cite{Milos+09}.  However, modestly exceeding the Eddington rate is
certainly plausible in theory: density inhomogeneities can allow
radiation to leak out of low density regions while most of the
accreting matter can be contained in high density regions.  For
example, magnetized radiation dominated accretion disks are subject to
a ``photon bubble'' instability that nonlinearly appears to lead to
strong density inhomogeneities (e.g. \cite{Begelman02}). Nevertheless,
observations have so far not revealed systems that sustain
super--Eddington accretion for extended periods; it would then still
have to be explained why the $z\approx6$ quasar BHs have this unique
behaviour.

\begin{figure}
  \includegraphics[height=.4\textheight]{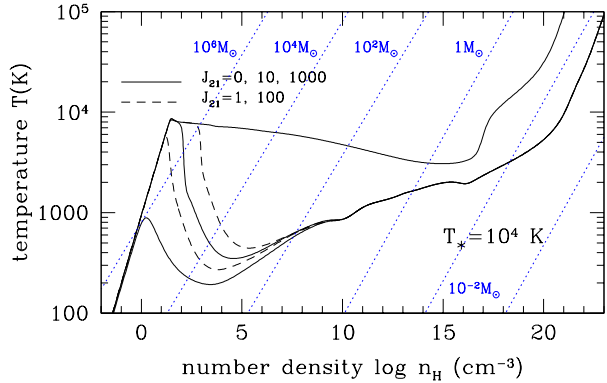}
  \caption{Temperature evolution of a metal-free cloud, irradiated by
    a strong UV flux.  The models solve for the chemical and thermal
    evolution, but assume a pre--imposed density evolution, based on
    the spherical collapse model.  Various cases are shown, with UV
    intensities at the Lyman limit of $J_{21}=0, 1, 10, 100$ and
    $10^{3}$, in the usual units of $10^{-21} {\rm
    erg~cm^{-2}~sr^{-1}~s^{-1}~Hz^{-1}}$ (solid and dashed curves; see
    the legend in the panel). Each blue dotted line corresponds to a
    different constant Jeans mass.  The gas is heated adiabatically
    until a density of $n\approx 10^0-10^2~{\rm cm^{-3}}$, at which
    ${\rm H_2}$--cooling becomes efficient and cools the gas to a few
    $\times 100$ K.  However, there exists a critical flux, with a
    value between $J_{21}=10^2$ and $10^3$, above which ${\rm
    H_2}$--cooling is disabled (adapted from ref.\cite{OSH08}).}
\label{fig:jcritonezone}
\end{figure}

\section{Growth by Rapid Direct Collapse}
\label{sec:directcollapse}

An appealing alternative idea is to produce, say, a $10^5~{\rm
M_\odot}$ SMBH ``directly'' -- i.e. much faster than this would take
under Eddington--limited accretion from a stellar seed.  This would
clearly be helpful to explain the high--redshift SMBHs, and many
authors (listed in the Introduction) proposed this may be possible
using metal--free gas in relatively massive ($\gsim 10^8~\msun$) dark
matter halos at redshift $z\gsim 10$, with virial temperatures $T_{\rm
vir}\gsim 10^4$K.

The gas that cools and collapses in these halos must avoid
fragmentation, shed angular momentum efficiently, and collapse
rapidly.  These conditions are unlikely to be met, unless the gas
remains ``warm'', i.e. at temperatures $T_{\rm vir}\sim 10^4$K.  In
recent numerical simulations, Shang et al. \cite{SBH10} found that the
gas in such halos, when collapsing in isolation, forms ${\rm H_2}$
efficiently, and cools to temperatures of $T \sim 300$ K.  Although no
fragmentation was seen, the gas is expected to ultimately fragment on
smaller scales that have not yet been resolved \cite{TAO09}.  More
importantly, even if fragmentation was avoided, the cold gas was found
to flow inward at low velocities, near the sound speed of $\sim
2-3~{\rm km~s^{-1}}$, with a correspondingly low accretion rate of
$\sim 0.01~{\rm M_\odot~yr^{-1}}$. This results in conditions nearly
identical to those in the cores of lower-mass minihalos; extensive
ultra--high resolution simulations had concluded that the gas then
forms a single $\sim 100~{\rm M_\odot}$ star \cite{ABN02,BCL02,YOH08}
or perhaps a massive binary \cite{TAO09}, rather than a supermassive
star or BH.

There have been at least three different ideas on how to avoid ${\rm
H_2}$--cooling and keep the gas warm.  One is for the gas to
``linger'' for a sufficiently long time at $10^4$K that it collapses
to a SMBH, even before ${\rm H_2}$ has a chance to reduce the
temperature.  For a sufficiently high space-- and column--density of
neutral hydrogen, the absorption of trapped Lyman $\alpha$ photons can
be followed by collisional de--excitation, rather than the resonant
scattering of the Lyman $\alpha$ photon, effectively trapping much of
the cooling radiation. This could lead to such lingering and to SMBH
formation -- analogous to opacity--limited fragmentation in colder gas
in the context of star formation \cite{SS06,SSG10}.

\begin{figure}
  \includegraphics[height=.3\textheight]{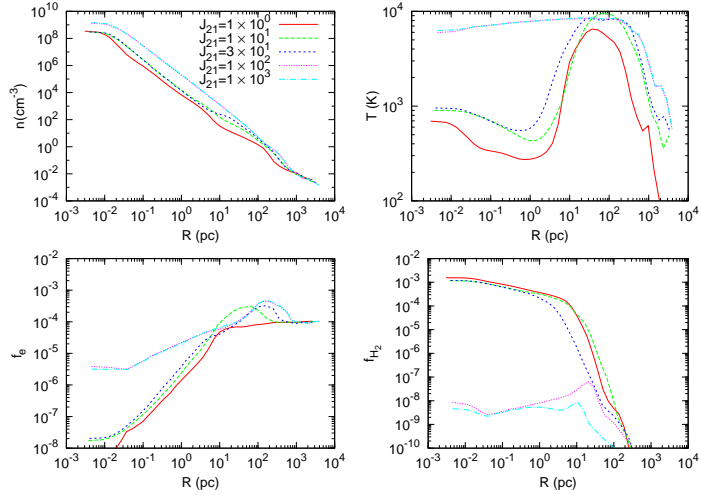}
  \caption{The results of adaptive mesh refinement (AMR) simulations
    of a primordial halo, with a total mass of a few $\times 10^7~{\rm
      M_\odot}$, collapsing at redshift $z\approx 10-15$, exposed to
    various UV background fluxes.  The four panels show snapshots of
    the spherically averaged profile of the particle number density,
    gas temperature, ${\rm e^{-}}$ fraction and ${\rm H_2}$ fraction
    at the time of the collapse of the core for several different
    values of the UV background intensity $J_{21}$, as labeled.  The
    existence of a critical flux, here with a value between
    $J_{21}=30$ and $10^2$, above which ${\rm H_2}$--cooling is
    disabled, is evident (adapted from ref.\cite{SBH10}).}
\label{fig:jcritsim}
\end{figure}

Alternatively, ${\rm H_2}$--cooling may be disabled if the gas is
exposed to an intense UV flux $J$, either directly photo--dissociating
${\rm H_2}$ (in the Lyman--Werner bands near a photon energy of $\sim
12$ eV) or photo--dissociating the intermediary ${\rm H^-}$ (at photon
energies $\gsim 0.76$ eV).  Requiring the photo-dissociation timescale,
$t_{\rm diss}\propto J^{-1}$, to be shorter than the ${\rm
H_2}$--formation timescale, $t_{\rm form}\propto \rho^{-1}$,
generically yields a critical flux that increases linearly with
density, $J^{\rm crit} \propto \rho$.  Since the gas in halos with
$T_{\rm vir}\gsim 10^4$K can cool via atomic Lyman $\alpha$ radiation
and lose pressure support, it inevitably collapses further. As a
result, in these halos, the critical flux is high, $J^{\rm
crit}\approx10^{2}-10^{5}$, depending on the assumed spectral shape
(ref.~\cite{SBH10}; see also refs. \cite{Omukai01,BL03} who found
similar, but somewhat higher values).  The existence of this critical
flux is illustrated in Figure~\ref{fig:jcritonezone}, using a one-zone
model in which the density evolution is approximated by spherical
collapse.  Figure~\ref{fig:jcritsim} shows the radial structure of a
$10^8{\rm M_\odot}$ halo, at the time of its collapse, when
illuminated at various intensities, taken from three--dimensional
simulations with the AMR code Enzo.  These profiles clearly show
that when the UV flux exceeds a critical value, the core of the halo
is prevented from cooling to low temperatures.

\begin{figure}
  \includegraphics[height=.3\textheight]{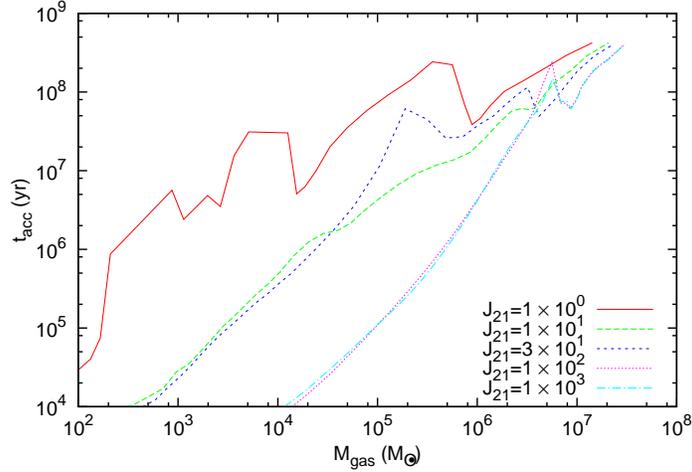} 
  \caption{The local accretion timescale $t_{\rm acc}$ as a function
    of the enclosed gas mass $M_{\rm gas}$, in the same halo depicted
    in Figure~\ref{fig:jcritsim}, illuminated with different
    intensities, as labeled.  In the halos exposed to a supercritical
    flux ($J_{21}=10^2$ and $10^3$), the mass accretion rate,
    $\dot{M}\approx 1~{\rm M_\odot\,yr^{-1}}$, is nearly $10^3$ times
    higher than in halos whose gas cools via ${\rm H_2}$ ($J_{21}\lsim
    10$).  At the center of the brightly illuminated halos, $\sim
    10^5 {\rm M_\odot}$ of gas accumulates within a Kelvin-Helmholtz
    time of $\sim 10^5$ yr, possibly leading to the
    formation of a SMBH with a comparable mass (adapted from
    ref.\cite{SBH10}).}
\label{fig:Macc}
\end{figure}

The 3D simulations also provide an estimate of the mass of the central
``object'' (star or SMBH) that ultimately forms at the core of the
halo, based on the measured profile of the mass accretion rate. This
is illustrated in Figure~\ref{fig:Macc}.  In particular, when the flux
exceeds the critical value, and the gas remains warm, the collapse is
significantly delayed.  However, when the gas ultimately does
collapse, it accretes toward the center at the sound speed
($c_s\approx 10$km/s), leading to a mass accretion rate of
$\dot{M}\approx 1{\rm M_\odot}\,{\rm yr^{-1}}$. This is much higher
than in the case of cold ($c_s\sim 1$ km/s) gas in halos with
efficient ${\rm H_2}$ cooling (the simulations reveal $\dot{M}\propto
c_s^3$, as expected in self--gravitating gas).

Importantly, the critical flux is high -- likely significantly
exceeding the expected level of the cosmic UV background at high
redshifts.  Therefore, only a small subset of all $T_{\rm vir}\gsim
10^4$K halos, which have unusually close and bright neighbors, may see
a sufficiently high flux.  However, given the strong clustering of
early halos, there is a sufficient number of these close halo pairs to
account for the abundance of the $z=6$ quasars \cite{Dijkstra+08}. A
more significant challenge to this idea is that in order to avoid
fragmentation, the gas in these halos must also remain essentially
free of any metals and dust \cite{OSH08}.  This requirement could be
difficult to reconcile with the presence of a nearby, luminous
galaxies.

\section{Alternative Models}
\label{sec:others}

Since both of the ``standard'' scenarios discussed above require some
optimistic assumptions, it is interesting to consider some more exotic
possibilities.

It is commonly believed that the magnetic fields permeating galaxies
such as the Milky Way arose by the amplification of a much weaker
large--scale seed field.  Weak primordial magnetic fields, with
strengths of up to $\sim$ 1nG, can be produced in phase transitions in
the early universe, during inflation, or during the electroweak or QCD
phase transitions.  It has recently been shown that such a primordial
magnetic field could produce a variant of the ``direct collapse''
scenario \cite{SHP10}. In particular, if the field is tangled, then
ambipolar diffusion will provide an efficient new mechanism to heat
the gas as it collapses in protogalactic halos. If the field has a
strength above $\mid B\mid\gsim 3.6$ (comoving) nG, the collapsing gas
is kept warm ($T\sim 10^4$ K) until it reaches the critical density
$n_{\rm crit}\approx10^3 {\rm cm^{-3}}$ at which the roto--vibrational
states of ${\rm H_2}$ approach local thermodynamic equilibrium.  ${\rm
H_2}$--cooling then remains inefficient, and the gas temperature stays
near $\sim 10^4$K, even as it continues to collapse to higher
densities. The critical magnetic field strength required to
permanently suppress ${\rm H_2}$--cooling is somewhat higher than
upper limit of $\sim 2$nG from the cosmic microwave background
(CMB). However, it can be realized in the rare $\gsim(2-3)\sigma$
regions of the spatially fluctuating $B$--field; these regions contain
a sufficient number of halos to account for the $z\approx6$ quasar BHs
\footnote{Because of the high magnetic Jeans mass, the magnetic
pressure has significant dynamical effects, and can prevent gas
collapse in halos with masses up to $M\gsim {\rm few}\times
10^{10}{\rm M_\odot}$. These are $\sim$100 times more massive than the
DM halos in the ``usual'' direct collapse models.}

Another ``exotic'' idea is that the first Pop III stars may be powered
by heating by dark matter annihilation, rather than by nuclear fusion
\cite{Spolyar+08}.  Weakly interacting massive particles (WIMPs) can
be such a heat source, as long as they reach sufficiently high density
inside the first stars, and if the annihilation products are trapped
inside the star.  Several authors have studied the impact of this
additional heating mechanism on the structure and evolution of such
``dark stars'' \cite{Spolyar+09,Iocco+08,Yoon+08,Taoso+08,Umeda+09,
Spolyar+09,Freese+10,Ripamonti+10}.  In particular, these stars can
live much longer than ``normal'' Pop III stars, and because their
radiation is soft, they can continue to accrete gas, as long as the
dark matter heating persists, and grow to masses of up to $\sim
10^5{\rm M_\odot}$ \cite{Umeda+09,Freese+10}.  These stars are bright,
and should be detectable directly by {\it JWST} \cite{Freese+10}.

\begin{figure}
  \includegraphics[height=.3\textheight]{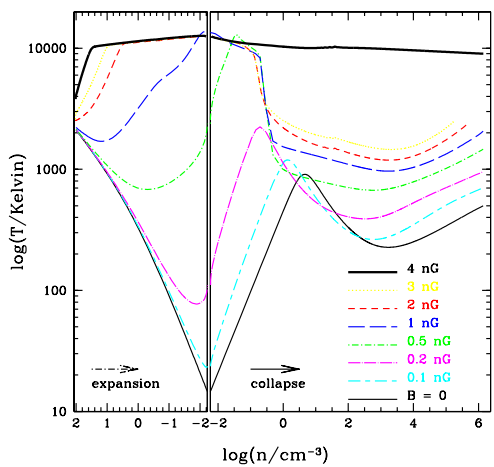}  
  \caption{The temperature evolution of a patch of the intergalactic
    medium is shown as it initially expands and then turns around and
    collapses to high density.  The different curves correspond to
    different values of the assumed primordial magnetic field, as
    labeled.  The gas evolves from the left to the right on this
    figure.  The left panel shows the expanding phase, starting from
    an initial density of $\approx 100~{\rm cm^{-3}}$ (corresponding
    to the mean density at redshift $z\simeq 800$) and ending at the
    turnaround just below $n=10^{-2}~{\rm cm^{-3}}$. The right panel
    follows the subsequent temperature evolution in the collapsing
    phase.  The figure shows the existence of a critical magnetic
    field, with a value between $B=3$ and $4$ nG, above which ${\rm
    H_2}$--cooling is disabled, and the gas temperature always remains
    near $10^4$K (adapted from ref.\cite{SHP10}). }
\label{fig:Bfield}
\end{figure}

\section{Observational Prospects}
\label{sec:observations}

In order to distinguish between the various proposed models discussed
above, observations must be able to detect SMBHs with masses below
$10^6~{\rm M_\odot}$ beyond redshift $z\approx 10$.  This requirement
is satisfied by {\it JWST} (with a sensitivity of $\sim10$ nJy at
near-IR wavelengths) and also in the radio by EVLA and, ultimately,
the proposed instrument SKA (with sensitivities of 1-10 $\mu$Jy at
1-10 GHz).  These IR and radio data should also be able to provide at
least a crude measure of their masses, but obtaining redshifts will be
challenging.

The assembly of the earliest SMBHs will ultimately be best probed by
the {\it LISA} satellite, which could detect $M<10^4~{\rm M_\odot}$
black holes beyond redshift $z \sim 20$, with a signal--to--noise ratio of
S/N > 30 or better \cite{Baker+07}. This should allow at least a crude
determination of the masses and distances to these low--mass SMBHs,
and to directly distinguish between the various proposed scenarios .


\begin{theacknowledgments}
  I thank the organizers for their kind invitation and for putting
  together a very stimulating conference.  I would also like to thank
  my recent collaborators, especially Cien Shang, Taka Tanaka, Mark
  Dijkstra, and Kazuyuki Omukai, whose work was especially emphasized
  here. The work described here was supported in part by the National
  Science Foundation and by the Pol\'anyi Program of the Hungarian
  National Office for Research and Technology (NKTH).
\end{theacknowledgments}

\bibliographystyle{aipproc}   

\begin{thebibliography}{51}
\expandafter\ifx\csname natexlab\endcsname\relax\def\natexlab#1{#1}\fi
\providecommand{\enquote}[1]{``#1''}
\expandafter\ifx\csname url\endcsname\relax
  \def\url#1{\texttt{#1}}\fi
\expandafter\ifx\csname urlprefix\endcsname\relax\def\urlprefix{URL }\fi
\providecommand{\eprint}[2][]{\url{#2}}

\bibitem[{Fan}(2006)]{Fanreview06}
X.~{Fan}, \emph{New Astronomy Review} \textbf{50}, 665--671 (2006).

\bibitem[{Abel} et~al.(2002)]{ABN02}
T.~{Abel}, G.~L. {Bryan}, and M.~L. {Norman}, \emph{Science} \textbf{295},
  93--98 (2002), \eprint{arXiv:astro-ph/0112088}.

\bibitem[{Bromm} et~al.(2002)]{BCL02}
V.~{Bromm}, P.~S. {Coppi}, and R.~B. {Larson}, \emph{\apj} \textbf{564}, 23--51
  (2002), \eprint{arXiv:astro-ph/0102503}.

\bibitem[{Yoshida} et~al.(2008)]{YOH08}
N.~{Yoshida}, K.~{Omukai}, and L.~{Hernquist}, \emph{Science} \textbf{321},
  669-- (2008), \eprint{0807.4928}.

\bibitem[{Heger} et~al.(2003)]{Heger+03}
A.~{Heger}, C.~L. {Fryer}, S.~E. {Woosley}, N.~{Langer}, and D.~H. {Hartmann},
  \emph{\apj} \textbf{591}, 288--300 (2003), \eprint{arXiv:astro-ph/0212469}.

\bibitem[{Haiman} and {Loeb}(2001)]{HL01}
Z.~{Haiman}, and A.~{Loeb}, \emph{\apj} \textbf{552}, 459--463 (2001),
  \eprint{arXiv:astro-ph/0011529}.

\bibitem[{Haiman}(2004)]{Haiman04}
Z.~{Haiman}, \emph{\apj} \textbf{613}, 36--40 (2004),
  \eprint{arXiv:astro-ph/0404196}.

\bibitem[{Yoo} and {Miralda-Escud{\'e}}(2004)]{YM04}
J.~{Yoo}, and J.~{Miralda-Escud{\'e}}, \emph{\apjl} \textbf{614}, L25--L28
  (2004), \eprint{arXiv:astro-ph/0406217}.

\bibitem[{Bromley} et~al.(2004)]{BSF04}
J.~M. {Bromley}, R.~S. {Somerville}, and A.~C. {Fabian}, \emph{\mnras}
  \textbf{350}, 456--472 (2004), \eprint{arXiv:astro-ph/0311008}.

\bibitem[{Shapiro}(2005)]{Shapiro05}
S.~L. {Shapiro}, \emph{\apj} \textbf{620}, 59--68 (2005),
  \eprint{arXiv:astro-ph/0411156}.

\bibitem[{Volonteri} and {Rees}(2006)]{VR06}
M.~{Volonteri}, and M.~J. {Rees}, \emph{\apj} \textbf{650}, 669--678 (2006),
  \eprint{arXiv:astro-ph/0607093}.

\bibitem[{Pelupessy} et~al.(2007)]{PDC07}
F.~I. {Pelupessy}, T.~{Di Matteo}, and B.~{Ciardi}, \emph{\apj} \textbf{665},
  107--119 (2007), \eprint{arXiv:astro-ph/0703773}.

\bibitem[{Li} et~al.(2007)]{Li+07}
Y.~{Li}, L.~{Hernquist}, B.~{Robertson}, T.~J. {Cox}, P.~F. {Hopkins},
  V.~{Springel}, L.~{Gao}, T.~{Di Matteo}, A.~R. {Zentner}, A.~{Jenkins}, and
  N.~{Yoshida}, \emph{\apj} \textbf{665}, 187--208 (2007),
  \eprint{arXiv:astro-ph/0608190}.

\bibitem[{Sijacki} et~al.(2009)]{SSH09}
D.~{Sijacki}, V.~{Springel}, and M.~G. {Haehnelt}, \emph{\mnras} \textbf{400},
  100--122 (2009), \eprint{0905.1689}.

\bibitem[{Tanaka} and {Haiman}(2009)]{TH09}
T.~{Tanaka}, and Z.~{Haiman}, \emph{\apj} \textbf{696}, 1798--1822 (2009),
  \eprint{0807.4702}.

\bibitem[{Oh} and {Haiman}(2002)]{OH02}
S.~P. {Oh}, and Z.~{Haiman}, \emph{\apj} \textbf{569}, 558--572 (2002),
  \eprint{arXiv:astro-ph/0108071}.

\bibitem[{Bromm} and {Loeb}(2003)]{BL03}
V.~{Bromm}, and A.~{Loeb}, \emph{\apj} \textbf{596}, 34--46 (2003),
  \eprint{arXiv:astro-ph/0212400}.

\bibitem[{Koushiappas} et~al.(2004)]{KBD04}
S.~M. {Koushiappas}, J.~S. {Bullock}, and A.~{Dekel}, \emph{\mnras}
  \textbf{354}, 292--304 (2004), \eprint{arXiv:astro-ph/0311487}.

\bibitem[{Lodato} and {Natarajan}(2006)]{LN06}
G.~{Lodato}, and P.~{Natarajan}, \emph{\mnras} \textbf{371}, 1813--1823 (2006),
  \eprint{arXiv:astro-ph/0606159}.

\bibitem[{Spaans} and {Silk}(2006)]{SS06}
M.~{Spaans}, and J.~{Silk}, \emph{\apj} \textbf{652}, 902--906 (2006),
  \eprint{arXiv:astro-ph/0601714}.

\bibitem[{Begelman} et~al.(2006)]{BVR06}
M.~C. {Begelman}, M.~{Volonteri}, and M.~J. {Rees}, \emph{\mnras} \textbf{370},
  289--298 (2006), \eprint{arXiv:astro-ph/0602363}.

\bibitem[{Volonteri} et~al.(2008)]{VLN08}
M.~{Volonteri}, G.~{Lodato}, and P.~{Natarajan}, \emph{\mnras} \textbf{383},
  1079--1088 (2008), \eprint{0709.0529}.

\bibitem[{Wise} and {Abel}(2008)]{WA08}
J.~H. {Wise}, and T.~{Abel}, \emph{\apj} \textbf{685}, 40--56 (2008),
  \eprint{0710.3160}.

\bibitem[{Regan} and {Haehnelt}(2009)]{RH09b}
J.~A. {Regan}, and M.~G. {Haehnelt}, \emph{\mnras} \textbf{393}, 858--871
  (2009), \eprint{0810.0024}.

\bibitem[{Schleicher} et~al.(2010)]{SSG10}
D.~R.~G. {Schleicher}, M.~{Spaans}, and S.~C.~O. {Glover}, \emph{\apjl}
  \textbf{712}, L69--L72 (2010), \eprint{1002.2850}.

\bibitem[{Shang} et~al.(2010)]{SBH10}
C.~{Shang}, G.~L. {Bryan}, and Z.~{Haiman}, \emph{\mnras} \textbf{402},
  1249--1262 (2010), \eprint{0906.4773}.

\bibitem[{Volonteri} and {Rees}(2005)]{VR05}
M.~{Volonteri}, and M.~J. {Rees}, \emph{\apj} \textbf{633}, 624--629 (2005),
  \eprint{arXiv:astro-ph/0506040}.

\bibitem[{Omukai} et~al.(2008)]{OSH08}
K.~{Omukai}, R.~{Schneider}, and Z.~{Haiman}, \emph{\apj} \textbf{686},
  801--814 (2008), \eprint{0804.3141}.

\bibitem[{Devecchi} and {Volonteri}(2009)]{DV09}
B.~{Devecchi}, and M.~{Volonteri}, \emph{\apj} \textbf{694}, 302--313 (2009),
  \eprint{0810.1057}.

\bibitem[{Begelman} et~al.(2008)]{BRA08}
M.~C. {Begelman}, E.~M. {Rossi}, and P.~J. {Armitage}, \emph{\mnras}
  \textbf{387}, 1649--1659 (2008), \eprint{0711.4078}.

\bibitem[{Wyithe} and {Loeb}(2003)]{WL03}
J.~S.~B. {Wyithe}, and A.~{Loeb}, \emph{\apj} \textbf{595}, 614--623 (2003),
  \eprint{arXiv:astro-ph/0304156}.

\bibitem[{Barkana} et~al.(2001)]{BHO01}
R.~{Barkana}, Z.~{Haiman}, and J.~P. {Ostriker}, \emph{\apj} \textbf{558},
  482--496 (2001), \eprint{arXiv:astro-ph/0102304}.

\bibitem[{Haiman} et~al.(1996)]{HTL96}
Z.~{Haiman}, A.~A. {Thoul}, and A.~{Loeb}, \emph{\apj} \textbf{464}, 523--+
  (1996), \eprint{arXiv:astro-ph/9507111}.

\bibitem[{Tegmark} et~al.(1997)]{Tegmark+97}
M.~{Tegmark}, J.~{Silk}, M.~J. {Rees}, A.~{Blanchard}, T.~{Abel}, and
  F.~{Palla}, \emph{\apj} \textbf{474}, 1--+ (1997),
  \eprint{arXiv:astro-ph/9603007}.

\bibitem[{Shankar}(2009)]{Shankarreview}
F.~{Shankar}, \emph{New Astronomy Review} \textbf{53}, 57--77 (2009),
  \eprint{0907.5213}.

\bibitem[{Alvarez} et~al.(2009)]{AWA09}
M.~A. {Alvarez}, J.~H. {Wise}, and T.~{Abel}, \emph{\apjl} \textbf{701},
  L133--L137 (2009), \eprint{0811.0820}.

\bibitem[{Milosavljevi{\'c}} et~al.(2009)]{Milos+09}
M.~{Milosavljevi{\'c}}, V.~{Bromm}, S.~M. {Couch}, and S.~P. {Oh}, \emph{\apj}
  \textbf{698}, 766--780 (2009), \eprint{0809.2404}.

\bibitem[{Begelman}(2002)]{Begelman02}
M.~C. {Begelman}, \emph{\apjl} \textbf{568}, L97--L100 (2002),
  \eprint{arXiv:astro-ph/0203030}.

\bibitem[{Turk} et~al.(2009)]{TAO09}
M.~J. {Turk}, T.~{Abel}, and B.~{O'Shea}, \emph{Science} \textbf{325}, 601--
  (2009), \eprint{0907.2919}.

\bibitem[{Omukai}(2001)]{Omukai01}
K.~{Omukai}, \emph{\apj} \textbf{546}, 635--651 (2001),
  \eprint{arXiv:astro-ph/0011446}.

\bibitem[{Dijkstra} et~al.(2008)]{Dijkstra+08}
M.~{Dijkstra}, Z.~{Haiman}, A.~{Mesinger}, and J.~S.~B. {Wyithe}, \emph{\mnras}
  \textbf{391}, 1961--1972 (2008), \eprint{0810.0014}.

\bibitem[{Sethi} et~al.(2010)]{SHP10}
S.~K. {Sethi}, Z.~{Haiman}, and K.~{Pandey}, \emph{ArXiv e-prints}  (2010),
  \eprint{1005.2942}.

\bibitem[Spolyar et~al.(2008)]{Spolyar+08}
D.~Spolyar, K.~Freese, and P.~Gondolo, \emph{Phys. Rev. Lett.} \textbf{100},
  051101 (2008).

\bibitem[{Spolyar} et~al.(2009)]{Spolyar+09}
D.~{Spolyar}, P.~{Bodenheimer}, K.~{Freese}, and P.~{Gondolo}, \emph{\apj}
  \textbf{705}, 1031--1042 (2009), \eprint{0903.3070}.

\bibitem[{Iocco} et~al.(2008)]{Iocco+08}
F.~{Iocco}, A.~{Bressan}, E.~{Ripamonti}, R.~{Schneider}, A.~{Ferrara}, and
  P.~{Marigo}, \emph{\mnras} \textbf{390}, 1655--1669 (2008),
  \eprint{0805.4016}.

\bibitem[{Yoon} et~al.(2008)]{Yoon+08}
S.~{Yoon}, F.~{Iocco}, and S.~{Akiyama}, \emph{\apjl} \textbf{688}, L1--L4
  (2008), \eprint{0806.2662}.

\bibitem[{Taoso} et~al.(2008)]{Taoso+08}
M.~{Taoso}, G.~{Bertone}, G.~{Meynet}, and S.~{Ekstr{\"o}m}, \emph{\prd}
  \textbf{78}, 123510--+ (2008), \eprint{0806.2681}.

\bibitem[{Umeda} et~al.(2009)]{Umeda+09}
H.~{Umeda}, N.~{Yoshida}, K.~{Nomoto}, S.~{Tsuruta}, M.~{Sasaki}, and
  T.~{Ohkubo}, \emph{Journal of Cosmology and Astro-Particle Physics}
  \textbf{8}, 24 (2009), \eprint{0908.0573}.

\bibitem[{Freese} et~al.(2010)]{Freese+10}
K.~{Freese}, C.~{Ilie}, D.~{Spolyar}, M.~{Valluri}, and P.~{Bodenheimer},
  \emph{\apj} \textbf{716}, 1397--1407 (2010), \eprint{1002.2233}.

\bibitem[{Ripamonti} et~al.(2010)]{Ripamonti+10}
E.~{Ripamonti}, F.~{Iocco}, A.~{Ferrara}, R.~{Schneider}, A.~{Bressan}, and
  P.~{Marigo}, \emph{\mnras} pp. 883--+ (2010), \eprint{1003.0676}.

\bibitem[{Baker} et~al.(2007)]{Baker+07}
J.~G. {Baker}, S.~T. {McWilliams}, J.~R. {van Meter}, J.~{Centrella},
  D.~{Choi}, B.~J. {Kelly}, and M.~{Koppitz}, \emph{\prd} \textbf{75},
  124024--+ (2007), \eprint{arXiv:gr-qc/0612117}.

\end{thebibliography}

\hyphenation{Post-Script Sprin-ger}

\end{document}